\documentclass{optica-article}

\journal{opticajournal} 

\articletype{Research Article}
\usepackage{lineno}
\usepackage{graphicx}
\usepackage{bbm}
\usepackage{bm}
\usepackage{hyperref}
\usepackage{makecell} 
\usepackage{enumerate}
\hypersetup{colorlinks=true,citecolor=blue,urlcolor=blue,linkcolor=blue}

\begin{document}

\title{Wavepacket interference of two photons through a beam splitter: from temporal entanglement to wavepacket shaping}

\author{Zhaohua Tian,\authormark{1} Qi Liu,\authormark{1,2} Yu Tian,\authormark{1,2} and Ying Gu\authormark{1,2,3,4,5,*}}

\address{\authormark{1}State Key Laboratory for Mesoscopic Physics, Department of Physics, Peking University Beijing 100871, China.\\
\authormark{2}Frontiers Science Center for Nano-optoelectronics \& Collaborative Innovation Center of Quantum Matter \& Beijing Academy of Quantum Information Sciences, Peking University, Beijing 100871, China\\
\authormark{3}Collaborative Innovation Center of Extreme Optics, Shanxi University, Taiyuan, Shanxi 030006, China\\
\authormark{4}Peking University Yangtze Delta Institute of Optoelectronics, Nantong 226010, China\\
\authormark{5}{Hefei National Laboratory, Hefei 230088, China}
}

\email{\authormark{*}ygu@pku.edu.cn} 


\begin{abstract*} 
Quantum interferences based on beam splitting are widely used for entanglement.  However, the quantitative measurement of the entanglement  in terms of temporal modes and wavepacket shaping facilitated by this entanglement remain unexplored. Here we analytically study the interference of two photons with different temporal shapes through a beam splitter (BS), then propose its application in temporal entanglement and shaping of photons.  The temporal entanglement  described by Von Neumann entropy is determined by the splitting ratio of BS and temporal indistinguishability of input photons. We found that maximum mode entanglement can be achieved with a 50/50 BS configuration, enabling the generation of a Bell state encoded in temporal modes, independent of the exact form of the input photons. Then, detecting one of the entangled photons at a specific time enables the probabilistic shaping of the other photon. This process can shape the exponentially decaying (ED) wavepacket into the ED sine shapes, which can be further shaped into Gaussian shapes with fidelity exceeding 99\%. The temporal entanglement and shaping of photons based on interference may solve the shape mismatch issues in large-scale optical quantum networks.
\end{abstract*}


\section{Introduction}
The non-classical interference processes among photons based on beam splitting are essential for both fundamental physics \cite{HOM1987PRL} and applications in quantum information processing \cite{KnillNature2001Linear,PieteRMP2007}. Perfect photon interference requires the photons to be perfectly indistinguishable, which is very challenging in experiment, especially for solid-state single photon emitters \cite{ MilosNP2016}. As a result, the interference processes of partially indistinguishable photons are receiving increasing attention \cite{OuPRA2006,OuIntJModPhysB2007, BarryPRL2013, ShchesnovichPRA2015, MaltePRA2015}. The effects of photon indistinguishability on output state probability and correlations have been systematically studied \cite{PhilipPRX2015,WalmsleyPRL2017, WalmsleyPRL2020}. Additionally, advancements in detection methods have enabled a thorough investigation of the temporal dynamics of multiphoton interference through time-resolved measurements, revealing much richer information. For example, the perfect NOON state can still be generated at specific moments even for detuned photons \cite{LegeroAPB2003,LaibacherPRL2015, RempePRL2004,PanPRL2018,AnthonyPRL2024}. Quantum beat patterns on correlation function can also be observed which are considered as signatures of entanglement \cite{MonroePRA2014,TammaPRA2018,PanPRL2014,JonathanPRAPP2022}. However, a comprehensive and quantitative measurement of the entanglement in terms of temporal modes is lacking.

The temporal shape of photons also plays a crucial role in the light-matter interactions \cite{PRA2010FanFullInversion,PRL2012FanStimu,Optica2019Alu,LiaoPRA2016}, deterministic quantum state transfer \cite{PRL1997CiracQST} and implementation of quantum logic gates \cite{PRL2020Dirk,PRApplied2020Zou}. 
Therefore, shaping the wavepacket of photons on demand has emerged as a pivotal aspect of quantum photonics. Current methods for shaping the photons include manipulating the emission of the quantum emitters dynamically \cite{PRL1997CiracQST} or passively \cite{PRApp2021Zhaohua}, direct phase modulation \cite{FanTimeReverModulators,FanTimeReverGaugePoten}, spectral or temporal filtering \cite{KimPRA2008,KonradOE2020}, and manipulating one of the entangled photon pairs \cite{LeuchsPRA2017,LeuchsPRA2017GenericMethod}. Nevertheless, the temporal shaping based on entanglement generated from photon wavepacket interference via beam splitting remains unexplored.

\par In this research, we study the interference of two photon wavepackets with different temporal shapes through a beam splitter (BS). The two initially  separable photons become entangled temporally after passing through the BS. Employing Schmidt decomposition approach to analyze the entanglement entropy, the measure of entanglement for output state are obtained. The entanglement degree is determined by  the photon temporal indistinguishability and splitting ratio of BS.  By utilizing this entanglement and detecting one of the entangled photons at a specific time, temporal shaping of the other photon can be probabilistically achieved in a heralded way. There are two exact examples: (1) the exponentially decaying (ED) wavepacket can be shaped into the ED sine shapes and (2) the ED sine can be further shaped into Gaussian shapes with fidelity higher than 99\%. The temporal entanglement and shaping of photons based on interference may address the shape mismatch issues in complex large-scale optical quantum networks.
\par The paper is organized as follows. In Sec. \ref{sec:II}, we give the analytical expression of the output state. A quantitative and comprehensive description of the temporal entanglement of the output state is given in Sec. \ref{sec:III}. Then the photon temporal shaping scheme based on beam splitting and two exact examples are demonstrated in \ref{sec:IV}. Finally, we conclude in Sec. \ref{sec:V}.

\section{Analytical expression  of the output state}
\label{sec:II}
\begin{figure}[!h]
	\centering
	\includegraphics*[width=60mm]{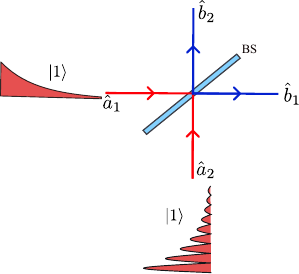}
	\caption{Schematic for the wavepacket interference of two photons through a beam splitter.}
	\label{fig-1}
\end{figure}
As shown in Fig. \ref{fig-1}, we consider the interference of two photons with different temporal shapes through a two-port BS.
The input state is a two-mode Fock state
\begin{equation}
	\begin{aligned}
		|\psi_\text{in}\rangle=\int f_1\left(\tau_1\right) \hat{a}_1^{\dagger}\left(\tau_1\right)d \tau_1 \int f_2\left(\tau_2\right)     \hat{a}_2^{\dagger}\left(\tau_2\right) d \tau_2|0,0\rangle,
	\end{aligned}
\end{equation}
where $f_{n}(\tau_{n})$ is the temporal shape of the single photon wavepacket in the $n$-th input port of BS and satisfies the normalization condition $\int |f_{n}(\tau)|^2d\tau=1$, and $\hat{a}^{\dagger}_{n}(\tau)$ ($\hat{a}_{n}(\tau)$) is the continuous time creation (annihilation) operator \cite{BlowPRA1990ContOperator}. The linear relationship between the operators in the input and output ports (denoted by $\hat{b}_{n}(\tau)$) can be described by the scattering matrix $\left(\begin{array}{c}
	\hat{b}_{1}(\tau)\\\hat{b}_{2}(\tau)
\end{array}\right)=\left(\begin{array}{cc}
	t & r \\
	-r & t
\end{array}\right)\left(\begin{array}{c}
	\hat{a}_{1}(\tau)\\\hat{a}_{2}(\tau)
\end{array}\right)$, where $t,r$ are the transmission and reflection coefficients of BS. The commutation relations of these continuous mode operators are   $[\hat{a}_{n}(\tau_{1}),\hat{a}^{\dagger}_{n}(\tau_{2})]=\delta(\tau_{1}-\tau_{2}),\quad [\hat{b}_{n}(\tau_{1}),\hat{b}^{\dagger}_{n}(\tau_{2})]=\delta(\tau_{1}-\tau_{2})$.  For convenience, we set $t$ and $r$ to be real numbers. For a lossless BS, $t^2 + r^2 = 1$.  
\par After passing through the BS, the output state could be expressed as
\begin{equation}
	|\psi_\text{out}\rangle =\sqrt{P_{0,2}}|0,2\rangle+\sqrt{P_{1,1}}|1,1\rangle+\sqrt{P_{2,0}}|2,0\rangle.
\end{equation}
Here $P_{m,2-m}$  denotes the probabilities of different outcomes which can be expressed as
\begin{subequations}\label{eq:Probability}
	\begin{align}
		\begin{split}
			& P_{2,0}=P_{0,2}=t^2 r^2\left(1+|J|^2\right),
		\end{split}
		\\
		\begin{split}
			& P_{1,1}=t^4+r^4-2t^2 r^2|J|^2,
		\end{split}
	\end{align}
\end{subequations}
where $J=\int f_1(\tau ) f_2^{*}(\tau ) \, d\tau$ is the temporal indistinguishability factor between the two input photons and $0\le |J|\le 1$. $|J|=1$ indicates perfect photon indistinguishability and $|J|=0$ means the two photons are completely distinguishable.  Therefore, the probability of each outcome is solely determined by the photon indistinguishability factor and splitting ratio of BS. The components $|0,2\rangle,|1,1\rangle,|2,0\rangle$ are expressed as follows
\begin{subequations}\label{Eq:outputstatedetail}
	\begin{align}
		\begin{split}
			|0,2\rangle
			&\equiv \iint F_{0,2}\left(\tau_1, \tau_2\right) \frac{\hat{b}_{2}^{\dagger}\left(\tau_1\right) \hat{b}_{2}^{\dagger}\left(\tau_2\right)}{\sqrt{2}} d \tau_1 d \tau_2|0,0\rangle,	
		\end{split}
		\\
		\begin{split}
			|1,1\rangle
			&\equiv\iint F_{1,1}\left(\tau_1, \tau_2\right) \hat{b}_{1}^{\dagger}\left(\tau_1\right) \hat{b}_{2}^{\dagger}\left(\tau_2\right) d \tau_1 d \tau_2|0,0\rangle,
		\end{split}
		\\
		\begin{split}
			|2,0\rangle 
			&\equiv \iint F_{2,0}\left(\tau_1, \tau_2\right) \frac{\hat{b}_{1}^{\dagger}\left(\tau_1\right) \hat{b}_{1}^{\dagger}\left(\tau_2\right)}{\sqrt{2}} d \tau_1 d \tau_2|0,0\rangle,	
		\end{split}
	\end{align}
\end{subequations}
where $F_{2,0}(\tau_{1},\tau_{2})=-F_{0,2}(\tau_{1},\tau_{2})=rt[f_1\left(\tau_1\right) f_2\left(\tau_2\right)+f_2\left(\tau_1\right) f_1\left(\tau_2\right)]/\sqrt{P_{2,0}}$, and $F_{1,1}(\tau_{1},\tau_{2})=[t^2 f_1\left(\tau_1\right) \allowbreak f_2\left(\tau_2\right)-\allowbreak r^2 f_2\left(\tau_1\right) f_1\left(\tau_2\right)]/\sqrt{P_{1,1}}$. Here $F_{m,2-m}(\tau_{1},\tau_{2})$ denotes the temporal shape or wavefunction of the component $|m,2-m\rangle$, normalized by $\iint |F_{m,2-m}(\tau_{1},\tau_{2})|^2 d\tau_{1}d\tau_{2}=1$. Therefore, after passing through the BS, the wavefunction of different components of the output state includes the coherent superposition of the product of temporal shapes $f_{n}(\tau)$, whose inseprable suggests the presence of temporal entanglement. Such temporally entangled photon pairs are an important quantum resource, but their generation primarily relies  on various nonlinear processes. For instance, spatially separated entangled photon pairs, like the component $|1,1\rangle$, can be produced through a type-II spontaneous parametric down-conversion (SPDC) process \cite{PanPRL2014}. Meanwhile, temporally entangled photon pairs in the same spatial mode, like the component $|2,0\rangle$ or $|0,2\rangle$, can be generated through a collinear type-0 or type-I SPDC process \cite{KulikPRA2001SPDC,FedorovPRA2008,MarcusOE2021},  or by coupling uncorrelated photons with a nonlinear system \cite{FanPRL2007,StolyarovPRA2019,LiaoPRA2010}.  The information of the interference-generated temporal entanglement is embedded within the analytical expressions of the wavefunction, which will be explored in depth in the following section.

\section{Temporal entanglement of the output state}
\label{sec:III}
It can be seen from  Eq. \eqref{Eq:outputstatedetail} that the wavefunction of the output state components $|2,0\rangle,|1,1\rangle$, and $|0,2\rangle$  is inseprable, revealing that the two photons are entangled temporally. This implies that temporal entanglement can be generated through the wavepacket interference of two photons with different temporal shapes. Prior studies have primarily concentrated on measuring the joint spectral density (JSD) profile \cite{RempePRL2004,MittalPRL2019,PanPRL2018,LaibacherPRL2015,NamPRA2015}, specifically focusing on the entanglement of the outcome $|1,1\rangle$. 
Here we give a comprehensive and quantitative examination of the entanglement in terms of temporal modes  across all three outcomes $|2,0\rangle,|1,1\rangle$, and $|0,2\rangle$.

\par First, we present the analytical expression for quantifying the entanglement of each component of the output state. We employ the Von Neumann entropy $\mathcal{S}_{|m,2-m\rangle}$ as an exact measure of entanglement, which can be expressed in terms of Schmidt coefficients as
\begin{equation}\label{eq:Entropy}
	\begin{aligned}
		\mathcal{S}_{|1,1\rangle}&=-\sum_{n=\pm}|\lambda_{n}^{1}|^2\log_{2}|\lambda_{n}^{1}|^2,\\
		\mathcal{S}_{|2,0\rangle}&=\mathcal{S}_{|0,2\rangle}=-\sum_{n=\pm}|\lambda_{n}^{2}|^2\log_{2}|\lambda_{n}^{2}|^2.
	\end{aligned}
\end{equation}
The Schmidt coefficients $\lambda_{\pm}^{m}$ satisfy the normalization condition $\sum_{n=\pm}|\lambda_{n}^{m}|^2=1$ and are obtained by performing Schmidt decomposition on the wavefunction of the output state components (see Supplement 1). The analytical expressions of Schmidt coefficients $\lambda_{\pm}^{m}$ are

\begin{subequations}\label{eq:SchmidtCoe}
	\begin{align}
		\begin{split}
			& |\lambda^{1}_{\pm}|^2=\frac{\mathcal{P}^2+\frac{1}{2} \mathcal{J}^2\left(1-\mathcal{P}^2\right) \pm \mathcal{P} \sqrt{|J|^2 \mathcal{J}^2(\mathcal{P}-1)^2+\left(\mathcal{P}-\mathcal{J}^2(\mathcal{P}-1)\right)^2}}{2(\mathcal{P}^2+\frac{1}{2} \mathcal{J}^2\left(1-\mathcal{P}^2\right))},		
		\end{split}
		\\
		\begin{split}
			|\lambda^{2}_{\pm}|^2=\frac{(1\pm |J|)^2}{2(1+|J|^2)},
		\end{split}
	\end{align}
\end{subequations}
where $\mathcal{J}=\sqrt{1-|J|^2}$, and $\mathcal{P}=t^2-r^2$  is defined to characterize the photon pathways indistinguishability. If $|\mathcal{P}|=1$, then $t^2=1$ or $t^2=0$, the photon pathways are completely distinguishable. Perfect indistinguishable photon pathways need $\mathcal{P}=0$ which means that $t^2=r^2=1/2$.  The Von Neumann entropy $\mathcal{S}$ is in the range $0\le\mathcal{S}\le 1$. $\mathcal{S}>0$ indicates the two photons are entangled and  $\mathcal{S}=0$ means quantum state  is separable. The maximum entanglement $\mathcal{S}=1$ is achieved when the two Schmidt mode coefficients are equal $|\lambda_{+}^{m}|^2=|\lambda_{-}^{m}|^2$, signifying the realization of the two-photon Bell state encoded in the temporal mode.  The correctness of the analytical results for the Schmidt decomposition  presented above has been verified through numerical methods (see Supplement 1).  According to Eqs. \eqref{eq:Entropy} and \eqref{eq:SchmidtCoe}, the entanglement of the output state is solely determined by the photon temporal indistinguishability $J$ and the transmission ratio $t^2$ of the BS. Additionally, the entanglement of different components varies: the entanglement of the components $|2,0\rangle$ and $|0,2\rangle$ are equal which are only related to the photon temporal indistinguishability $J$, but the entanglement of the component $|1,1\rangle$ is determined by both the photon indistinguishability and the splitting ratio of BS. 

\begin{figure}[!t]
\centering
	\includegraphics*[width=86mm]{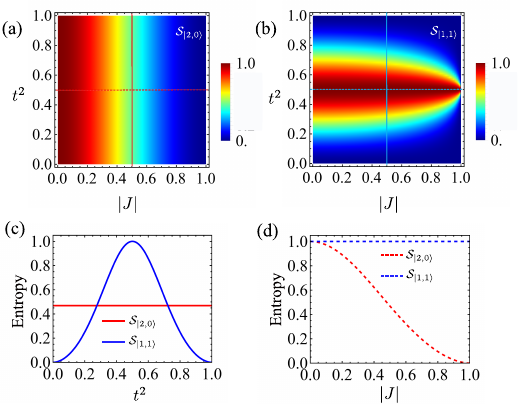}
	\caption{Von Neumann entropy of the output state for the outcome (a) $|2,0\rangle$ and (b) $|1,1\rangle$ as a function of photon indistinguishability and transmission of BS;  Slice of the Von Neumann entropy of the outcome  $|1,1\rangle$  (blue line) and $|2,0\rangle$ (red line)  (c)  as a function transmission of BS (solid line) when $|J|=0.5$ and (d) as a function photon indistinguishability (dashed line) when $t^2=0.5$.}
	\label{fig-2}
\end{figure}

\par We then systematically investigate the effect of photon indistinguishability $|J|$ and transmission $t^2$ of BS on the Von Neumann entropys. As depicted in Figs. \ref{fig-2}(a) and (b), a common feature is that an increase in the photon indistinguishability $|J|$ leads to a decrease in the temporal entanglement for both $|1,1\rangle$ and $|2,0\rangle$, except for the case where $t^2=0.5$.  Maximal entanglement is achieved for $|J|=0$, i.e., the two input photons are completely distinguishable.  In the conventional HOM interference, a perfect NOON state can only be generated when the two photons are completely indistinguishable \cite{HOM1987PRL}, while our quantitative results show that when examining entanglement in terms of temporal modes, entanglement can only be generated when the two interfering photons are partly distinguishable. It is important to emphasize that the entanglement of $|2,0\rangle$ can be ambiguous for $J=0$. Two photons are entangled in the Schmidt modes basis $\phi_{\pm}^{2}(\tau)$ but separable in the original temporal modes $f_{1}(\tau),f_{2}(\tau)$. This ambiguity arises due to the arbitrariness of (temporal) mode selections (see Supplement 1), which is very common for mode entanglement in optics \cite{FabreRMP2020Modes}. Furthermore,   as illustrated in Fig. \ref{fig-2} (c), the entanglement of  $|1,1\rangle$ can be modulated by adjusting the beam splitting ratio of BS. The maximal entanglement is achieved  when $t^2=0.5$ ($\mathcal{P}=0$), i.e., the photon pathways are completely indistinguishable. Conversely, the entanglement is zero when $t^2=1$ or $0$ ($|\mathcal{P}|=1$), revealing that completely distinguishable photon pathways result in no entanglement for the outcome $|1,1\rangle$. These quantitative results can serve as valuable guidance for manipulating temporal entanglement generated through interference.

\par 
We also discover an intriguing result:  the temporal entanglement of component $|1,1\rangle$ always remains maximal with a 50/50 BS configuration, independent of photon indistinguishability. As depicted in Fig. \ref{fig-2} (d), when $t^2=0.5$, as the increase of the photon temporal indistinguishability $|J|$, the entanglement of
$|1,1\rangle$ consistently remains at 1. In this case, it can be proven that, excepting for perfect indistinguishable photons ($|J|=1$ so that $P_{1,1}=0$), regardless of the exact form of the two input photons,  the component  $|1,1\rangle$ can always be considered as a Bell state encoded in the temporal Schmidt modes $|\phi_{\pm}^{1}\rangle$ (see Supplement 1) 
\begin{equation}\label{eq:11BellState}
	|1,1\rangle=\frac{1}{\sqrt{2}}(|\phi_{+}^{1}\rangle_{\hat{b}_{1}}|\phi_{-}^{1}\rangle_{\hat{b}_{2}}-|\phi_{-}^{1}\rangle_{\hat{b}_{1}}|\phi_{+}^{1}\rangle_{\hat{b}_{2}}),
\end{equation}
then $|\lambda_{-}^1|^2=|\lambda_{+}^1|^2=1/2$ and $\mathcal{S}_{|1,1\rangle}=1$. The specific form of the input photons $f_{n}(\tau)$ determines the shapes of the Schmitt modes $\phi_{\pm}^{1}(\tau)$ but does not change the superposition coefficients between them.  This is because the outcome $|1,1\rangle$  arises from two indistinguishable photon pathways: two photons are either simultaneously reflected or transmitted by the BS. When $t^2=0.5$, then $\mathcal{P}=0$, the two photon pathways are completely indistinguishable. This facilitates the creation of maximized entangled Bell state, the entanglement of which does not depend on specific temporal shapes. When $t^2 \neq 0.5$, the two photon pathways are not completely indistinguishable, resulting in a non-maximally entangled state. Consequently, the entanglement is influenced by the indistinguishability of the input photons.
\par  It should be additionally emphasized that the aforementioned orthogonal Schmidt modes are not just a mathematical decomposition. They can also be used to encode quantum information \cite{BrechtPRX2015} and can be spatially separated through the nonlinear process \cite{EcksteinOE2011QuantumPulseGate}, thereby converting into entangled states in spatial modes.
The quantitative results above provide us with a more profound understanding of the mechanism and properties of temporal entanglement generated through wavepacket interference of two photons. This temporal entanglement, as a significant quantum resource, holds potential applications in fundamental physics tests and quantum information processing. In the following section, we will demonstrate its application in temporal shaping of photons.

\section{Single-photon temporal shaping}
\label{sec:IV}
In this section, we will demonstrate that temporal entanglement  generated through wavepacket interference of two photons can be utilized for shaping. For temporally entangled photon pairs, shaping one photon can be achieved by manipulating or measuring the other photon. Firstly, we present the relationship between the temporal shapes of the wave packets before and after shaping. Subsequently, we provide two specific examples of wavepacket shaping based on this relationship.

\subsection{Temporal shaping principle}

\begin{figure}[!th]
\centering
	\includegraphics[width=135mm]{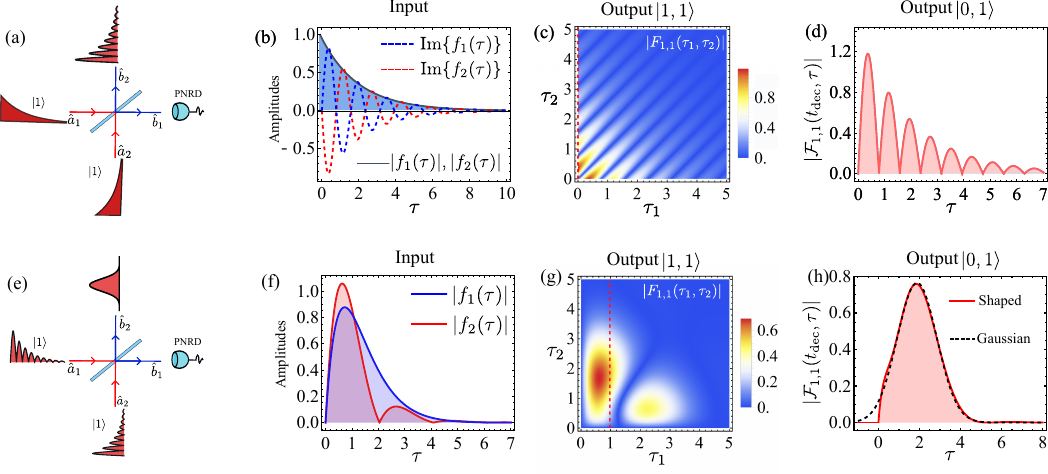}
	\caption{The temporal shaping scheme and examples.  (a)-(d) Single photon with ED sine shape generated via the interference of two ED shaped photons. (e)-(h) Single photon with Gaussian shape generated via the interference of two ED sine shaped photons. The detailed parameters are in the main text. (a), (e) Schematic for single photon temporal shaping based on the interference of two photon wavepackets and time-resolved measurement. A photon-number-resolved detector (PNRD) is placed at the output port 1 of BS.  (b), (f) The temporal shapes of the two input photons. (c), (g) The temporal shape of the component $|1,1\rangle$. The red dashed line denotes the required detection time instant. (d), (h) The temporal shape of resulting single photon state $|0,1\rangle$ in output port 2 after time-resolved detection.}
	\label{fig-3}
\end{figure}

We first present the shaping principle based on wavepacket interference and time-resolved measurement. Theoretically, the wavefunctions of all three possible outcomes are temporally inseparable and can all be utilized for temporal shaping. However, only the temporal entanglement of the outcome $|1,1\rangle$ can be controlled by the beam splitting ratio of BS, offering more degrees of freedom. So we will focus on the shaping based on the outcome $|1,1\rangle$ in the main text. The principle and example of temporal shaping based on the outcome $|2,0\rangle$ can be found in Supplement 1. Temporal shaping scheme based on the outcome $|1,1\rangle$ is depicted in Fig. \ref{fig-3}(a) (or Fig. \ref{fig-3}(e)), two photon wavepackets with different temporal shapes impinge on the BS and a photon-number-resolved detector (PNRD) is placed at the output port 1 of BS. By detecting only one photon at a specific time $t_{\text{dec}}$, the component $|1,1\rangle$ can be selected probabilistically. The resulting single photon state in output port 2 can be obtained 

\begin{equation}
		\begin{aligned}
			|0,1\rangle=\int \mathcal{F}_{1,1}(t_{\text{dec}},\tau)\hat{b}_{2}^{\dagger}(\tau)d\tau|0\rangle,
		\end{aligned}
\end{equation}
where  $\mathcal{F}_{1,1}(t_{\text{dec}},\tau)$ denotes the resulting temporal shape
\begin{equation}
	\mathcal{F}_{1,1}(t_{\text{dec}},\tau)=\frac{1}{\sqrt{\text{nor}}}\left(t^2 f_{1}(t_{\text{dec}})f_{2}(\tau)-r^2 f_{2}(t_{\text{dec}})f_{1}(\tau)\right).
\end{equation}
Here the normalization coefficient, $\text{nor}=\int | F_{1,1}(t_{\text{dec}},\tau)|^2 d\tau$, ensures the proper normalization of the state. The shape of the heralded single photon $\mathcal{F}_{1,1}(t_{\text{dec}},\tau)$ in output port 2 depends on the temporal shapes of two input photons, the transmission of BS,  and the time-resolved measurement outcome at a time instant $t_{\text{dec}}$. The joint control of these parameters tailors the temporal shape.

\subsection{Example: Shaping photon from ED shape into ED sine shape}

We now demonstrate shaping photon from ED shape into ED sine shape. As shown in \ref{fig-4}(a), 
two detuned ED shaped photons are incident on the entrance ports of a 50/50 BS. By detecting only a single photon at a desired time instant via a PNRD at output port 1 of the BS, single photons with an ED sine shape can be generated at output port 2. The temporal shapes of the two input photons can be expressed as $f_{1}(\tau)=\sqrt{\Gamma}e^{-i( \Delta\omega-i\Gamma)\tau/2}$,$f_{2}(\tau)=\sqrt{\Gamma}e^{-i(- \Delta\omega-i\Gamma)\tau/2}$, here $\Gamma,\Delta\omega$ denote the spectral width and detuning of two input photons, respectively. Following Eq. \eqref{Eq:outputstatedetail}, the wavefunction of the component $|1,1\rangle$ after passing the symmetric BS is $F_{1,1}(\tau_{1},\tau_{2})=i\sqrt{\frac{1}{P_{1,1}}}e^{-\Gamma(\tau_{1}+\tau_{2})/2}\sin(\Delta\omega (\tau_{2}-\tau_{1})/2).$
By detecting only one photon at time instant $t_{\text{dec}}=0$  at output port 1 of BS, the temporal shape of the resulting single photon  at output port 2 is
\begin{equation}
	\mathcal{F}_{1,1}(0,\tau)=i\sqrt{\frac{\Gamma\left( \Delta\omega^2+\Gamma^2\right)}{\Delta\omega^2/2}}e^{-\Gamma\tau/2}\sin(\Delta\omega\tau/2),
\end{equation}
which is the required ED sine shape. 
\par We offer a more intuitive demonstration of the shaping process described above with specific parameters. The temporal shapes of two input wavepackets with ED shapes are shown in Fig. \ref{fig-3} (b). The linewidth of the two photons are the same $\Gamma=1$ and the detuning is set to be $\Delta\omega=8$. After passing through the BS, the wavefunction of the component $|1,1\rangle$  is depicted in Fig. \ref{fig-3} (c), where  quantum beat pattern can be observed. By detecting only one photon at time instant $t_{\text{dec}}= 0$ at output port 1 of BS, as indicated by the red dashed lines in Fig. \ref{fig-3}(c), the resulting single photon at port 2 with required ED shape is shown in Fig. \ref{fig-3}(d).

\subsection{Example: Shaping photon from ED sine shape into Gaussian shape}
We further demonstrate shaping photon from ED sine shape into Gaussian shape, which is crucial for achieving high-fidelity deterministic optical quantum storage \cite{PRL1997CiracQST} and optical quantum logic gates \cite{PRL2020Dirk,PRApplied2020Zou}. As shown in Fig. \ref{fig-3}(e), two photons with different ED sine shapes impinge on the entrance ports of a BS. The temporal shape of the input photons can be described as $f_{n}(\tau)=\sqrt{\frac{\Gamma _n\left(4 \omega _n^2+\Gamma _n^2\right)}{2 \omega _n^2}} e^{  -\Gamma_n \tau/2} \sin \left( \omega_n \tau\right)$, where $\Gamma_{n}$ and $\omega_{n}$ represent the spectral width and resonant frequency, respectively.
The  Gaussian shape can be expressed as $f_{\text{Gau}}(\tau)=\sqrt{\frac{\Gamma_{\text{Gau}}}{\sqrt{\pi}}}e^{-\frac{1}{2}(\tau-\tau_{0})^2\Gamma_{\text{Gau}}^2}$, where $\Gamma_{\text{Gau}}$ and $\tau_{0}$ denote the spectral width and the delay time of photon wavepacket, respectively.  We aim for excellent indistinguishability between the interferometrically synthesized wavepacket and the Gaussian shape wavepacket. This translates into realizing the following optimization problem: 
\begin{equation}
	\text{Maximize: }	F_{\text{shaping}}(x)=\left|\int f_{\text{Gau}}^{*}(\tau)\mathcal{F}_{1,1}(t_{\text{dec}},\tau)d\tau\right|^2,
\end{equation}
here $F_{\text{shaping}}$ represents the shaping fidelity, defined as the square of the indistinguishability between the interferometrically synthesized wavepacket and the Gaussian wavepacket. In optimization,  the linewidth of the Gaussian shape  $\Gamma_{\text{Gau}}$ is fixed, and the parameters to be optimized are $x=\left\{\Gamma_{1},\Gamma_{2},\omega_{1},\omega_{2},t,\tau_0,t_{\text{dec}}\right\}$,
including the linewidth of the input photons $\Gamma_{n}$, resonant frequency $\omega_{n}$, transmission coefficients $t$ of BS, delay time of the shaped Gaussian shape $\tau_{0}$ and an appropriate detection time $t_{\text{dec}}$. 
\par 
Now, we intuitively demonstrate the process of obtaining high-fidelity Gaussian-shaped photons through shaping. We set the linewidth of the target Gaussian-shaped photons to be $\Gamma_{\text{Gau}}=1$. After optimization, the optimal parameters are $x=\left\{2.04,\allowbreak 2.60,-1.49\allowbreak,0.380,0.768,1.95,1.01\right\}$. The corresponding temporal shapes of the two input photons are shown in Fig.  \ref{fig-3}(f), both of which are ED sine shapes with different linewidths and resonant frequencies. The wavefunction of $|1,1\rangle$ after passing through the BS is shown in Fig. \ref{fig-3}(g). We can observe that the output wavefunction is entangled temporally. By detecting only one photon at  output port 1 of BS at time $t_{\text{dec}}=1.01$, as indicated by the red dashed lines in Fig. \ref{fig-3}(g), the resulting single photon temporal shape at output port 2 of BS is drawn in Fig. \ref{fig-3}(h). The target Gaussian shape is also shown in Fig. \ref{fig-3}(h) as a reference. We can observe that the temporal shape generated through interference is very similar to the target Gaussian shape. The corresponding shaping fidelity is 0.996. Compared to the shaping methods demonstrated in the Refs \cite{LeuchsPRA2017GenericMethod,LeuchsPRA2017}, our approach achieves high-fidelity Gaussian-shaped photons without the need for nonlinear process. 

\subsection{Discussions}
\par We briefly discuss the influence of time resolution of the detector on the shaping. In the above two examples, we have assumed that detectors could precisely detect photons at specific moments $t_{\text{dec}}$. However, in actual detection processes, the resolution of detectors is always limited. We can only confirm the detection of photons within a certain time interval $(t_{\text{dec}}-t_{\text{R}}/2,t_{\text{dec}}+t_{\text{R}}/2)$, where $t_{\text{R}}$ denotes the detection time resolution of the detector. We define the temporal width of the target ED sine (Gaussian) shaped photon as $\tau_{\text{width}}=1/\Gamma$  ($\tau_{\text{width}}=1/\Gamma_{\text{Gau}}$). The effects of detection time resolution $t_{\text{R}}$ on success rate and fidelity of shaping are demonstrated in suppplemnt 1.  The results show that the limited resolved time of detectors could increase the success rate but slightly decrease the fidelity of shaping. Nevertheless, the high fidelity remains nearlly unaffected for $t_{\text{R}}/\tau_{\text{width}}<0.1$, which is achievble for current detection technology  \cite{AnthonyPRL2024}.  Depending on the experimental goals, detection resolution can be flexibly selected to balance the trade-off between shaping success rate and fidelity.
\par The temporal shaping scheme discussed above provides new insights into wavepacket transformation in quantum information processing. Shaping photons on demand at various stages of optical quantum information processing is crucial. Failure to do so may result in temporal shape mismatch issues, which can affect photon interference and light-matter interactions, thereby further impacting the corresponding quantum information processes. In the two examples we have presented the conversion of commonly encountered single photon temporal shapes can be achieved, offering a solution to mitigate the impact of temporal shape mismatch issues in complex large-scale optical quantum networks. In principle, based on the wavepacket interference of two photons, more temporal shaping functionalities can be realized, thereby facilitating the adaptation to various task requirements in quantum information.

\section{Summary}
\label{sec:V}
In summary, we have studied the interference of two photon wavepackets with different temporal shapes through BS. We have employed the Von Neumann entropy to quantitatively describe the entanglement of the output state and found analytically that the entanglement is directly determined by the temporal indistinguishability of photons and the splitting ratio of BS.  When examining entanglement from the standpoint of temporal modes, entanglement can only be generated when the two interfering photons are partly distinguishable. Maximum entanglement of the component $|1,1\rangle$ can be achieved with a 50/50 BS configuration, enabling the generation of a Bell state encoded in temporal modes, independent of the exact form of the input photons. In a Schmitd mode basis, completely distinguishable input photons will maximize the entanglement of the outcomes $|2,0\rangle$ and $|0,2\rangle$. We have further demonstrated that the temporal shaping of a single photon can be achieved probabilistically by detecting one of the entangled photons. Our work reveals the crucial roles of photon temporal and photon pathway indistinguishability in the generation of temporal entanglement. The proposed scheme of single-photon temporal shaping also offers a solution to the issues of shape mismatch in complex large-scale optical quantum networks.
\begin{backmatter}
\bmsection{Acknowledgments}
We thank Fengxiao Sun and Shuheng Liu for fruitful discussions. 

\bmsection{Funding}
This work is supported by the National Natural Science Foundation of China under Grants No. 11974032 and by the Innovation Program for Quantum Science and Technology under Grant No. 2021ZD0301500.

\bmsection{Disclosures}
The authors declare no conflicts of interest.

\bmsection{Supplemental document}
See Supplement 1 for supporting content. 

\end{backmatter}




\end{document}